# Room temperature 9μm photodetectors and GHz heterodyne receivers

Daniele Palaferri[1], Yanko Todorov[1], Azzurra Bigioli[1], Alireza Mottaghizadeh[1], Djamal Gacemi[1], Allegra Calabrese[1], Angela Vasanelli[1], Lianhe Li[2], A. Giles Davies[2], Edmund H. Linfield[2], Filippos Kapsalidis[3], Mattias Beck[3], Jérôme Faist[3] and Carlo Sirtori[1]

[1] Laboratoire Matériaux et Phénomènes Quantiques, Université Paris Diderot, Sorbonne Paris Cité, CNRS-UMS 7162, 75013 Paris, France
[2] School of Electronic and Electrical Engineering, University of Leeds, Leeds LS2 9JT, United Kingdom
[3] ETH Zurich, Institute of Quantum Electronics, Auguste-Piccard-Hof 1, Zurich 8093, Switzerland

**Room temperature operation is mandatory for any optoelectronics technology which aims to provide low-cost compact systems for widespread applications. In recent years, an important technological effort in this direction has been made in bolometric detection for thermal imaging[1], which has delivered relatively high sensitivity and video rate performance (~ 60 Hz). However, room temperature operation is still beyond reach for semiconductor photodetectors in the 8–12 μm wavelength band[2], and all developments for applications such as imaging, environmental remote sensing and laser-based free-space communication[3-5] have therefore had to be realised at low temperatures. For these devices, high sensitivity and high speed have never been compatible with high temperature operation[6,7]. Here, we show that a 9 μm quantum well infrared photodetector[8], implemented in a metamaterial made of subwavelength metallic resonators[9-12], has strongly enhanced performances up to room temperature. This occurs because the photonic collection area is increased with respect to the electrical area for each resonator, thus significantly reducing the dark current of the device[13]. Furthermore, we show that our photonic architecture overcomes intrinsic limitations of the material, such as the drop of the electronic drift velocity with temperature[14,15], which constrains conventional geometries at cryogenic operation[6]. Finally, the reduced physical area of the device and its increased responsivity allows us, for the first time, to take advantage of the intrinsic high frequency response of the quantum detector[7] at room temperature. By beating two quantum cascade lasers[16] we have measured the heterodyne signal at high frequencies up to 4 GHz.**

An important intrinsic property of inter-subband (ISB) quantum well infrared photodetectors (QWIPs) based on III-V semiconductor materials that has not yet been exploited is the very short lifetime of the excited carriers. The typical lifetime is of the order of one picosecond[7], which leads to two important consequences: the detector frequency response can reach up to 100 GHz, and the saturation intensity is extremely high ($10^7$ W/cm$^2$)[17]. These figures are ideal for a heterodyne detection scheme where a powerful local oscillator (LO) can drive a strong photocurrent, higher than the detector dark current, that can coherently mix with a signal shifted in frequency with respect to the LO. Notably, these unique properties are unobtainable in infrared inter-band detectors based on mercury-cadmium-telluride (MCT) alloys, which have a much longer carrier lifetime and therefore an intrinsic lower speed response[2,18]. Yet, the performance of all photonic detectors is limited by the high dark current that originates from thermal emission of electrons from the wells, and rises exponentially with temperature, thus imposing cryogenic operation (∼ 80 K) for high sensitivity measurements. Although highly doped (∼$1\times10^{12}$ cm$^{-2}$) 10 μm QWIPs have been observed to operate up room temperature, tens of mW incident power from a $CO_2$ laser was required to measure the signal[19,20].

In the present work, we show that this intrinsic limitation in QWIP detectors can be overcome through use of a photonic metamaterial. We are able to calibrate our detector at room temperature using a black body emitting only hundreds of nW, orders of magnitude smaller than that required previously. To date, room temperature performance with values comparable to those that we report here has only been demonstrated in the 3–5 μm wavelength range, using quantum cascade detectors (QCDs)[21-23] and MCT standard detectors[24].

The photonic metamaterial structure is shown in Fig. **1a**. The GaAs/AlGaAs QWIP[8] contains $N_{qw}$ = 5 quantum wells absorbing at 8.9 μm wavelength (139 meV) that has been designed according to an optimized bound-to-continuum structure from ref. **7**. The absorbing region is inserted in an array of double-metal patch resonators[9-12], which provide sub-wavelength electric field confinement and act as antennas. The resonant wavelength is fixed by the patch size *s* through the expression $\lambda = 2sn_{eff}$, where $n_{eff}$ = 3.3 is the effective index[9]. As shown in the reflectivity measurement in Fig. **1b**, the cavity mode is in close resonance with the peak responsivity of the detector.

In our structure, the microcavity increases the device responsivity by a local field enhancement in the thin semiconductor absorber[10], while the antenna effect extends the photon collection area of the detector, $A_{coll}$, making it much larger than the electrical area $\sigma = s^2$ of the device[13]. As the detector photocurrent is proportional to $A_{coll}$, while the dark current is proportional to $\sigma$, for the same number of collected photons there is therefore a substantial reduction of the dark current that results in a net increase of the detector operating temperature.

Besides the collection area $A_{coll}$, which defines the absorption cross section per patch resonator, another crucial parameter is the contrast $C$ of the reflectivity resonance shown in Fig. **1b**. This parameter quantifies the fraction of the incident photon flux absorbed collectively by the array and should be as close as possible to 1. As shown in Fig. **1c**, the contrast can be adjusted by changing the array periodicity $p$[10]. Optimal detector responsivity is obtained at the *critical coupling point*, $C$ = 1, where all incident radiation is coupled into the array. The collection area per patch is related to the contrast according to the expression $A_{coll} = Cp^2\xi$, where the factor $\xi$ = 0.7 takes into account the polarizing effect of the connecting wires (see S.I.)[13]. From the data in Fig. **1c**, the critical coupling is obtained with a period $p$ = 3.3 µm, which corresponds to a collection area $A_{coll}$ = 7.5 µm², four times larger than the electrical area $\sigma$ = 1.7 µm² of the patch.

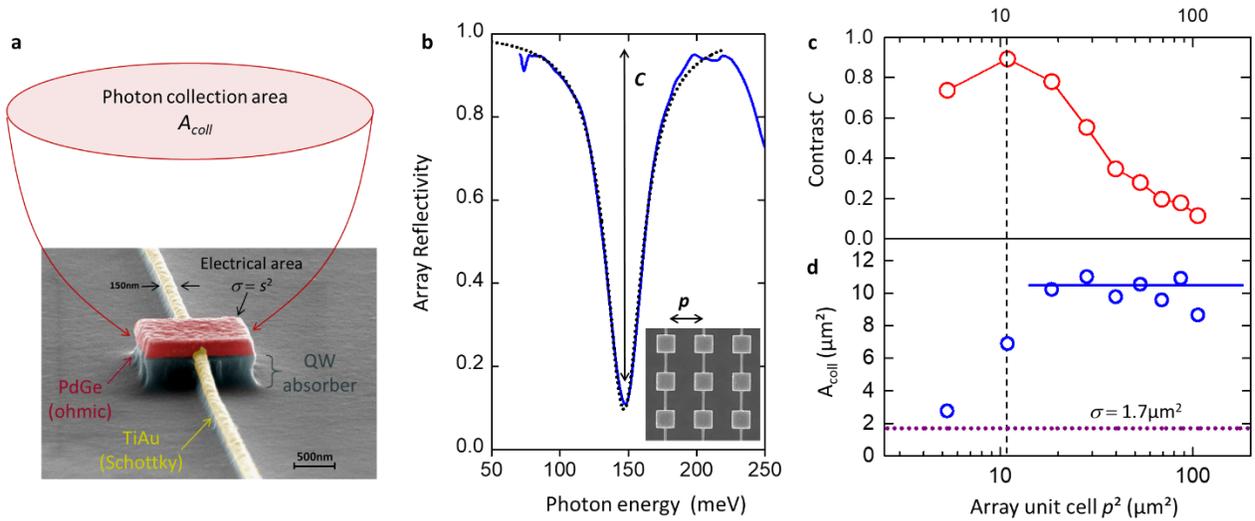

**Fig. 1 Device concept: a**, Double-metal antenna-coupled microcavity realized by e-beam lithography with ohmic alloy (PdGeTiAu) contacts and connected by 150 nm thin wires isolated by Schottky barriers (Ti/Au). The active region contains a QWIP structure (386 nm) with five QWs Si-doped at n=7×10$^{11}$ cm$^{-2}$. The pixel size of the array is 50 µm. This metamaterial structure allows photons to be collected from a collection area $A_{coll}$ that is much larger than the electrical surface area σ. **b**, reflectivity spectrum (blue curve) of an array with $s$=1.30 µm and $p$=3.30 µm polarized along the TM$_{100}$ mode (perpendicular to the thin wires): the dashed line is a Lorentzian fit yielding a contrast C = 1−R$_{dip}$ ~ 90%. The size of the patch cavity has been chosen to resonate with the ISB electronic transition E$_{12}$ ~ 139 meV. **c**, contrast and collection area A$_{coll}$=CΣ as function of the array unit cell area Σ=$p^2$. The collection area saturates for large unit cell periods, as expected from the theory in Ref. **13**. The contrast is optimal for arrays with a period $p$ =3.3 µm.

Notice that the device processing has been optimized in order to generate current solely under the metallic square patches and not below the 150 nm wide leads connecting them. To this end we have realised ohmic contacts between the patches and the underlying semiconductor layers using PdGeTiAu annealed alloy, while a Schottky barrier, made by depositing TiAu, prevents vertical current between the metallic wire and the semiconductor. Moreover, all cavities are connected to an external wire-bonding pad insulated by an 800-nm-thick $Si_3N_4$ layer (S.I.). Thanks to all these precautions the conductive area is reduced to the sum of the areas of all the patch resonators, which prevents additional dark current from flowing across the device.

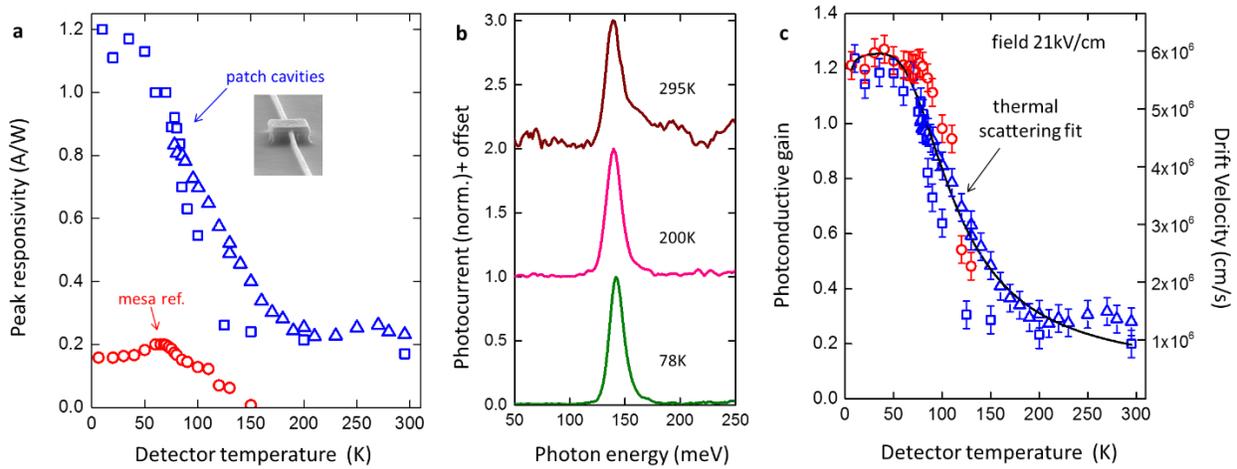

**Fig. 2 Responsivities & Spectra. a**, peak responsivity of QWIP devices fabricated in 200 µm diameter mesa (circles), and into patch resonator arrays with $s$=1.35 µm (squares) and $s$=1.30 µm (triangles), both with pixel size of 50 µm. The responsivities were measured with a calibrated 1000°C blackbody source as a function of detector heatsink temperature. **b**, normalized photocurrent spectra of the array with $s$=1.30 µm at 78 K, 200 K and 295 K. **c**, photoconductive gain and electronic drift of the three devices presented in **2a** as a function of temperature: the data are shown at a bias voltage 0.5 V corresponding to a field of 21 kV/cm. The drift velocity is obtained using a QW capture time of 5 ps (see ref. **7** and S.I.).

In order to quantify the detector performance, we have compared the detector array with a reference device, here referred to as "mesa", where the same absorbing region is processed into 200 µm diameter circular mesa and light is coupled in through the 45°-polished substrate edge[25]. The mesa reference provides the intrinsic photo-response of the detector (see S. I.). In Fig. **2a** we compare the peak responsivities for the two configurations, obtained with a calibrated black body source at 1000°C (see Methods for more details). The mesa device could be characterized only up to 150 K, as the photo-current becomes undetectable at higher temperatures. The array detectors show a seven-fold enhancement of the responsivity at low temperatures. Most remarkably, the responsivity could be characterized up to room temperature, where the measured responsivity (0.2 A/W) is comparable with the best responsivity for the mesa device measured at around 50 K. We were thus able to record photo-current spectra up to room temperature, Fig. **2b**, which is, to our knowledge, the first type of such measurement with a QWIP operating in the 9 µm band using a thermal source.

By quantifying carefully the number of photons absorbed in each geometry (Methods), we were also able to extract the photoconductive gain *g* for each structure (Fig. **2c**). We recall that the gain provides the number of electrons circulating per photon absorbed in the QWs[26,27], and is an intrinsic property of the detector absorbing region. All our devices show the same values of the gain as a function of temperature, irrespective of their fabrication geometry, which proves that the material properties are identical for the two structures. Following Ref.**7**, the photoconductive gain is proportional to the electron drift velocity in the AlGaAs barriers and its temperature dependence is linked to microscopic scattering processes in polar materials[14,15]. Our results fit well the temperature dependence of the drift velocity described on ref. **14**. The derived low temperature value of the drift velocity is of the order of 6×10$^6$ cm/s as expected at an electric field of 20 kV/cm for an Al concentration in the range 20–30%[28]. These results account for the temperature drop of the responsivity observed in Fig. **2a**. Above 200 K, the gain acquires an almost constant value *g* = 0.25 – 0.2, of the order of 1/$N_{qw}$. This implies that photoexcited electrons can only *travel* from one well to the next adjacent well, as the mean free path of the electrons is now shorter than the distance between two wells. Very interestingly, in this limit, it clearly appears that a detector based on a single QW would be advantageous at high temperatures. These results illustrate how our devices give access to the high temperature physics of quantum detectors, a unique regime unexplored so far.

The best assessment of detector performance is the background-limited specific detectivity[26] $D_B^* = \frac{R\sqrt{A_{det}}}{\sqrt{4egI}}$ plotted in Fig. **3a** for the mesa reference and for the patch devices. The experimental results are compared with our model that describes the impact of the photonic design on the detectivity as a function of the temperature[13]. For clarity, in Fig. **3b** we provide the ratio between the detectivities. At low temperature, we observe an enhancement of only a factor of two. Here, the dark current is negligible and the main source of noise is the background photocurrent induced by the 300 K black body of the environment. In this regime higher responsivity means also higher background noise, and the detectivity enhancement scales with the square root of the responsivities ratio i.e. ($R_{array}/R_{mesa}$)$^{1/2}$= 2.6. The situation is totally different at high temperature, where the dark current is the dominant contribution to the noise. In this case the detectivity enhancement is

$$R_{array}/R_{mesa} \, (A_{coll}/\sigma)^{1/2} \sim 14,$$

and the actual performance of the arrays at 300 K is equivalent to the performance of the mesa reference at 150 K, doubling the temperature of operation. This is a significant improvement, well beyond that is predictable from the low temperature operation. Our device concept therefore takes advantage of both the responsivity enhancement, and the strong suppression of the dark current owing to the antenna effect, as expressed by the factor ($A_{coll}/\sigma$)$^{1/2}$. As explained in Ref. **13**, the combination of the microcavity and the antenna effect thus slows down the decrease of the detectivity with temperature, pushing the detector operation to much higher temperatures than expected.

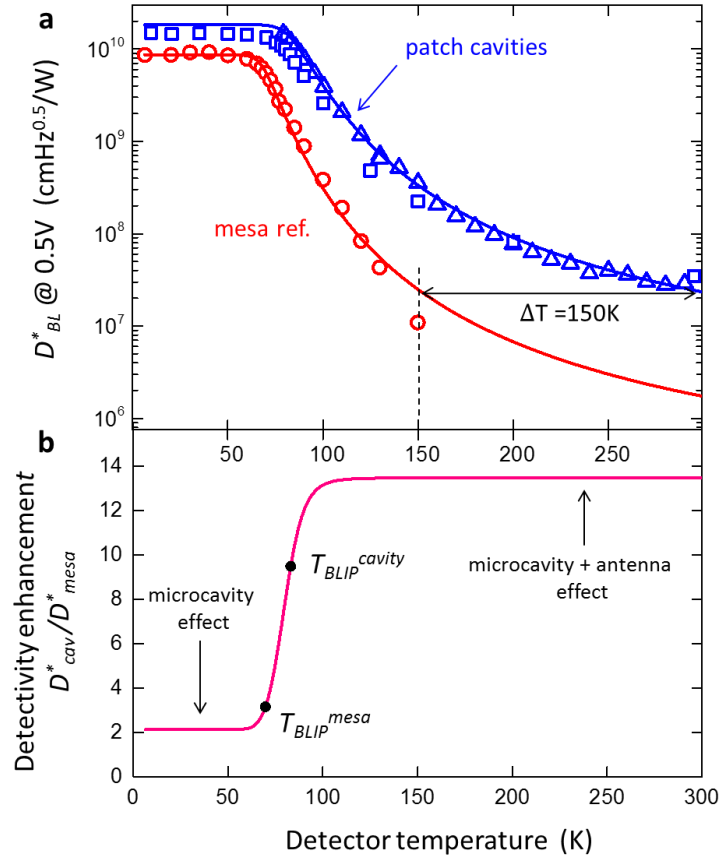

**Fig. 3 Background-limited detectivity. a**, background-limited specific detectivity (2π field of view) as a function of the temperature, in the case of the mesa (circles) and the arrays with s=1.30 µm (triangles) and s=1.35 µm (squares) at 0.5V. The red line is a fit of the mesa experimental data to $d(T)=d_0/[1+d_1 T \exp(-E_{act}/k_B T)]^{1/2}$ where $d_0$ and $d_0$ are fit parameters, $E_{act}$ = 120meV is the activation energy and $k_B$ is the Boltzmann constant. The blue curve is the model described in ref. **13**, expressing the performances of quantum detectors embedded in patch resonators. **b**, ratio between the detectivities in the two different detector geometries. Dots show the corresponding BLIP temperatures: $T_{BLIP}^{mesa}$ =70 K for the mesa and $T_{BLIP}^{cavity}$ =83 K for the patch cavity arrays.

By exploiting our photonic concepts we have achieved high temperature operation with relative high sensitivities. We now seek to benefit from the inherent very high frequency response together with the reduced electrical capacitance of our devices in order to use them as heterodyne receivers. In this case, by increasing the power of the local oscillator one may achieve the ultimate heterodyne sensitivity set only by the detector absorption coefficient. The heterodyne scheme allows the very fast frequency response of the QWIP detectors at room temperature, as well as their very high saturation intensity, to be fully exploited, paving the way to a new class of coherent quantum devices in the mid- and far-infrared spectral ranges.

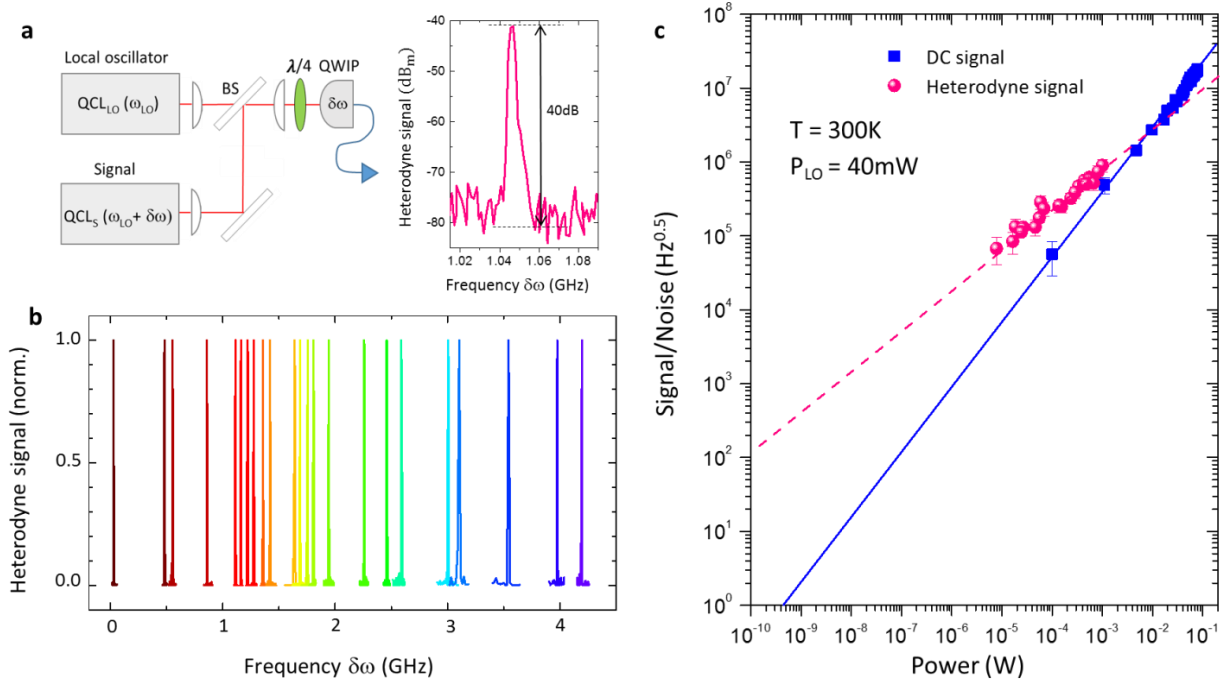

**Fig. 4 Tunable heterodyne experiment and results. a**, heterodyne arrangement involving two DFB QCLs and the room temperature QWIP in the cavity array geometry. A 40 dB heterodyne power spectrum is shown, acquired with a 1MHz resolution bandwidth of the spectrum analyzer. **b**, normalized heterodyne power signal (in linear scale) acquired using a spectrum analyzer. By slightly changing the current on the DFB lasers we can tune the positions of $\omega_S$ and $\omega_{LO}$ (see S.I.). **c**, log-log plot of the signal-to-noise ratio as function of the QCL power. The noise of the QWIP is calculated using the measured gain and dark current values at room temperature.

This realization is depicted in Fig. **4a**, where we show schematically the heterodyne arrangement that we used to probe our detector at room temperature. It consists of two DFB quantum cascade lasers (QCLs)[16] operating at $\lambda$ = 8.36 µm that are made collinear by a beam splitter before they impinge on the detector. The latter is connected via wire bonding to a high frequency coaxial cable that is connected to a spectrum analyser. Each laser has a linewidth of the order of one MHz when current and temperature are stabilised. In order to ensure that their individual frequencies are very close, one laser is kept at 293 K, while the second is kept at 254 K. When the detector is illuminated by both lasers a clear heterodyne signal appears on the spectrum analyser.

In Fig. **4a** we show a measurement at 1.06 GHz, with a 40 dB signal-to-noise ratio. We have measured heterodyne signals up to 4.2 GHz as it is illustrated in Fig. **4b**. Our bandwidth is presently limited by a strong impendence mismatch between the detector and the external circuit. In Fig. **4c** we report our first characterisation of the sensitivity of the heterodyne receiver at room temperature. The blue dots correspond to the D.C. saturation curve for the LO, while the orange curve is the heterodyne signal at 1 GHz as a function of the signal power. The straight line is a linear fit for the LO saturation curve. The saturation experiment shows that the detector responds linearly up to 78 mW (~ 3.1 kW/cm$^2$) of incident power. Moreover, the linear fit intercepts the 1 Hz integration band for a power of ~ 0.5 nW, in very good agreement with the measured detectivity at room temperature. As can be observed from Fig. **4c**, the heterodyne data are very well fitted with a square root dependence (dashed line) and can reach a signal-to-noise ratio of a few pW, for an integration time of the order of 10 ms. This clearly shows the strength of the heterodyne technique that let us envision sensitivity in the thermal region at λ = 9 μm which is unreachable with any other technique at room temperature. It has to be mentioned that the power of the LO in our experiment is still far from being the source of the highest current circulating into the device. We are in fact dominated by the dark current of the detector, $I_{dark}$ ~ 3.5 mA, while the local oscillator photocurrent is $I_{LO}$ ~ 0.5 mA. By increasing the LO power and/or decreasing the temperature of the detector by few tens of degrees using thermo-cooled elements, these detectors could reach the ultimate heterodyne detection limit, set by their absorption efficiency[7,13] and the relative intensity noise of the local oscillator[29].

In conclusion, we have demonstrated coherent heterodyne detection at 9 μm wavelength. By beating two single mode QC lasers with close frequencies we have produced a heterodyne signal up to 4.2 GHz which allows detecting phase and amplitude with unmatched sensitivity at room temperature. Moreover, this scheme could be also very efficient for the generation and synthesis of microwaves.

**References**

1.      Wood, R. A. Uncooled microbolometer infrared sensor arrays. In *Infrared detectors and emitters: materials and devices* **8** 149-175, edited by Springer, Boston https://doi.org/10.1007/978-1-4615-1607-1_6 (2001)

2.      Rogalski, A. Infrared detectors: status and trends. *Progress in quantum electronics* **27.2**, 59-210 http://dx.doi.org/10.1016/S0079-6727(02)00024-1 (2003)

3.      Gunapala, S. D. & Bandara, S. V. in Intersubband Transition in Quantum Wells: Physics and Device Applications I, *Semiconductors and Semimetals* **62**(4) 197–282, edited by H. C. Liu and F. Capasso Academic Press, San Diego (2000)

4.      Mizaikoff, B. Peer reviewed: Mid-ir fiber-optic sensors *Am. Chem. Soc.* **75** 258A http://pubs.acs.org/doi/pdf/10.1021/ac031340g  (2003)

5.      Martini, R., & E. A. Whittaker. Quantum cascade laser-based free space optical communications. *Journal of Optical and Fiber Communications Reports* **2**(4) 279-292 http://dx.doi.org/10.1007/s10297-005-0052-2  (2005)

6.      Henini, M. & Razeghi, M. *Handbook of infrared detection technologies* Elsevier, Oxford (2002)

7.      Schneider, H. & Liu, H.C. *Quantum Well Infrared Photodetectors Physics and Applications*, Springer, New York United States (2007)

8.      Levine, B. F., Choi, K. K., Bethea, C. G., Walker, J., & Malik, R. J. New 10 μm infrared detector using intersubband absorption in resonant tunneling GaAlAs superlattices. *Appl. Phys. Lett.* **50**(16) 1092-1094  http://dx.doi.org/10.1063/1.97928   (1987)

9.      Todorov, Y. *et al.* Optical properties of metal-dielectric-metal microcavities in the THz frequency range *Opt. Express* **18** 13886 https://doi.org/10.1364/OE.18.013886  (2010)

10.     C. Feuillet-Palma, Y. Todorov, A. Vasanelli, and C. Sirtori. Strong near field enhancement in THz nano-antenna arrays. *Sci. Rep.* **3** 1361 https://doi.org/10.1038/srep01361  (2013).

11.     Nga Chen, Y. *et al.* Antenna-coupled microcavities for enhanced infrared photo-detection. *Appl. Phys. Lett.* 104 031113 http://dx.doi.org/10.1063/1.4862750  (2014)

12.     Palaferri, D. *et al.* Patch antenna terahertz photodetectors. *Appl. Phys. Lett.* **106** 161102 http://dx.doi.org/10.1063/1.4918983  (2015)

13.     Palaferri, D. *et al.* Ultra-subwavelength resonators for high temperature high performance quantum detectors. *New J. Phys.* **18**(11) 113016 http://dx.doi.org/10.1088/1367-2630/18/11/113016  (2016)

14.     Grundmann, M. *Physics of Semiconductors* **11** Springer Berlin (2010).

**15.**    Howarth, D. J., & Sondheimer, E. H. The theory of electronic conduction in polar semi-conductors. *Proc. R. Soc. A*: *Mathematical, Physical and Engineering Sciences* **219**(1136) 53-74 http://dx.doi.org/10.1098/rspa.1953.0130  (1953)

**16**.    Faist, J. *et al.* Distributed feedback quantum cascade lasers. *Appl. Phys. Lett.* **70**(20) 2670-2672 http://dx.doi.org/10.1063/1.119208  (1997)

**17.**    Vodopyanov, K. L., Chazapis, V., Phillips, C. C., Sung, B., & Harris Jr, J. S. Intersubband absorption saturation study of narrow III-V multiple quantum wells in the spectral range *Semicond. Sci. Technol.* **12**(6) 708 https://doi.org/10.1088/0268-1242/12/6/011  (1997)

**18.**    Theocharous, E., Ishii, J. & Fox, N. P. A comparison of the performance of a photovoltaic HgCdTe detector with that of large area single pixel QWIPs for infrared radiometric applications *Infrared physics & technology* **46**(4) 309-322.  http://doi.org/10.1016/j.infrared.2004.05.002 (2005)

**19.**    Grant, P. D., Dudek, R., Buchanan, M., & Liu, H. C. Room-temperature heterodyne detection up to 110 GHz with a quantum-well infrared photodetector. *IEEE Photon. Technol. Lett.* **18**(21) 2218-2220 https://doi.org/10.1109/LPT.2006.884267  (2006)

**20.**    Hao, M. R., Yang, Y., Zhang, S., Shen, W. Z., Schneider, H., & Liu, H. C. (2014). Near-room-temperature photon-noise-limited quantum well infrared photodetector. *Laser & Photonics Reviews*, **8**(2), 297-302. https://doi.org/10.1002/lpor.201300147 (2014)

**21.**    Graf, M., Hoyler, N., Giovannini, M., Faist, J., & Hofstetter, D. InP-based quantum cascade detectors in the mid-infrared. *Appl. Phys. Lett. 88*(24), 241118 http://dx.doi.org/10.1063/1.2210088  (2006)

**22.**    Hofstetter, D. *et al.* Mid-infrared quantum cascade detectors for applications in spectroscopy and pyrometry. *Appl. Phys. B* **100**(2) 313-320 http://dx.doi.org/10.1007/s00340-010-3965-2  (2010)

**23.**    Hinds, S. *et al.* Near-Room-Temperature Mid-Infrared Quantum Well Photodetector. *Adv. Mater.* **23**(46) 5536-5539  http://dx.doi.org/10.1002/adma.201103372   (2011)

**24**.    Piotrowski, J., Galus, W., & Grudzien, M. Near room-temperature IR photo-detectors. *Infrared Physics* **31**(1) 1-48 https://doi.org/10.1016/0020-0891(91)90037-G  (1991)

**25.** Helm, M. The basic physics of intersubband transitions. *Semiconductors and semimetals* **62** 1-99 edited by H. C. Liu and F. Capasso, Academic Press, San Diego (2000).

**26.** Rosencher, E. & Vinter, B. *Optoelectronics* Cambridge University Press (2004).

**27.** Liu, H.C. Photoconductive gain mechanism of quantum-well intersubband infrared detectors *Appl. Phys. Lett.* **60** 1507 http://dx.doi.org/10.1063/1.107286 (1992)

**28**. Hava, S., & Auslender, M. Velocity-field relation in GaAlAs versus alloy composition. *J. Appl. Phys.* **73**(11) 7431-7434 http://dx.doi.org/10.1063/1.353985 (1993)

**29.** Gensty, T., Elsäßer, W., & Mann, C. Intensity noise properties of quantum cascade lasers. Opt. Express **13**(6) 2032-2039 https://doi.org/10.1364/OPEX.13.002032 (2005)

**Methods**

**QWIP fabrication.** The QWIP structure is grown by MBE (molecular beam epitaxy). It consists of five GaAs QWs, each with a thickness $L_{QW}$ = 5.2 nm and each *n*-doped across the central 4 nm region with Si at a density of $N_d$ = 1.75x10$^{18}$ cm$^{-3}$. The QWs are separated by $Al_{25}Ga_{75}As$ barriers of thickness $L_b$ = 35 nm. At the top and bottom of this periodic structure GaAs contact layers are grown, with thicknesses $L_{c,top}$ = 100.0 nm and $L_{c,bottom}$ = 50.0 nm and doping $N_{d,top}$ = 4.0x10$^{18}$ cm$^{-3}$ and $N_{d,bottom}$ = 3.0x10$^{18}$ cm$^{-3}$, respectively. The double-metal structures are obtained through wafer-bonding on a GaAs host substrate using 500 nm gold layers, and by selectively etching down to an etch-stop $Al_{65}Ga_{35}As$ layer grown before the bottom contact. As shown in Fig. **1a**, the patch-antennae are connected by 150 nm thin metallic wires which are realized using electron-beam lithography (consecutive alignments allow different metallic alloy contacts). The final structure is obtained by ICP etching of the semiconductor region between the antennae. The 45° facet substrate-coupled geometry consists of a 200 μm diameter circular mesa, with annealed Pd/Ge/Ti/Au as a top contact and annealed Ni/Ge/Au/Ni/Au as a diffused bottom contact.

**Reflectivity and photocurrent analysis.** Reflectivity spectra and photocurrent spectra were obtained using a Bruker Vertex interferometer. Reflectivity measurements were performed at a 15° incident angle and at room temperature. For the photocurrent spectra, QWIP devices were mounted in a cryostat with an internal cooled metallic shield and a ZnSe optical window. Photocurrent and responsivity were measured using a blackbody source at 1000 °C, which was calibrated with an MCT detector. The source is focused onto the detector by two gold parabolic mirrors (f/1 and f/3), providing typical field of view of 60°. The photocurrent is measured with a lock-in technique using an optical chopper at 1059 Hz and a shunt resistance connected to the voltage input of a lock-in amplifier Stanford Research SR1830, without using pre-amplifiers.

**Heterodyne measurement.** The two beams from the QCLs are made collinear using f/0.5 germanium lenses and a beam splitter, and then focused onto the detector by a f/1.5 lens and a λ/4 waveplate to avoid optical feedback (Fig. **3a**). The two lasers are DC biased with a voltage supply and are mounted in two Janis cryostats to stabilize their temperatures using liquid nitrogen flow. The QWIP is polarized by a Keythley 2450 sourcemeter and the heterodyne signal is sent to the spectrum analyser Agilent E4407B using a bias tee. In this arrangement the QWIP detector is at room temperature, without using any cooling system.

**Acknowledgements**

We acknowledge financial support from the FP7 ITN NOTEDEV project (Grant. No. 607521), the ERC grant "ADEQUATE", the French National Research Agency (ANR-16-CE24-0020 Project "hoUDINi"), and the EPSRC (UK) projects "COTS" and "HYPERTERAHERTZ" (EP/J017671/1, EP/P021859/1). EHL and AGD acknowledge the Royal Society and the Wolfson Foundation, and thank Dr L. Chen for skilled support with the device processing.

**Author contributions**

D.P., Y.T. and C.S. conceived the experiments, designed the QWIP structure, analysed the data and wrote the manuscript. D.P. fabricated the QWIP devices and performed measurements and data analysis together with A.B. A.M. and D.G. helped with the heterodyne measurements. A.C. calibrated the blackbody for the responsivity measurements and helped with the characterization of the mesa device. A.V. helped with data analysis. L.L., A.G.D. and E.H.L. grew the QWIP structure and provided the wafer-bonding for the double-metal processing. F.K., M.B. and J.F. provided the DFB QCLs for the heterodyne experiment. All the work has been realised under the supervision of C.S.

**Competing financial interests**

The authors declare no competing financial interests.

# Supporting Information

# Room temperature 9μm photodetectors and GHz heterodyne receivers

**Device architecture of the QWIP patch antenna array**

Fig. S1 shows a SEM image of the quantum detector device made of our metamaterial photonic concept. The pixel of the device is 50x50μm². The external pad is connected to the array by the 150nm wires and is insulated from the bottom ground plane by a 800nm thick $Si_3N_4$ layer. The TiAu pad connects the device to the external circuit by wire bonding.

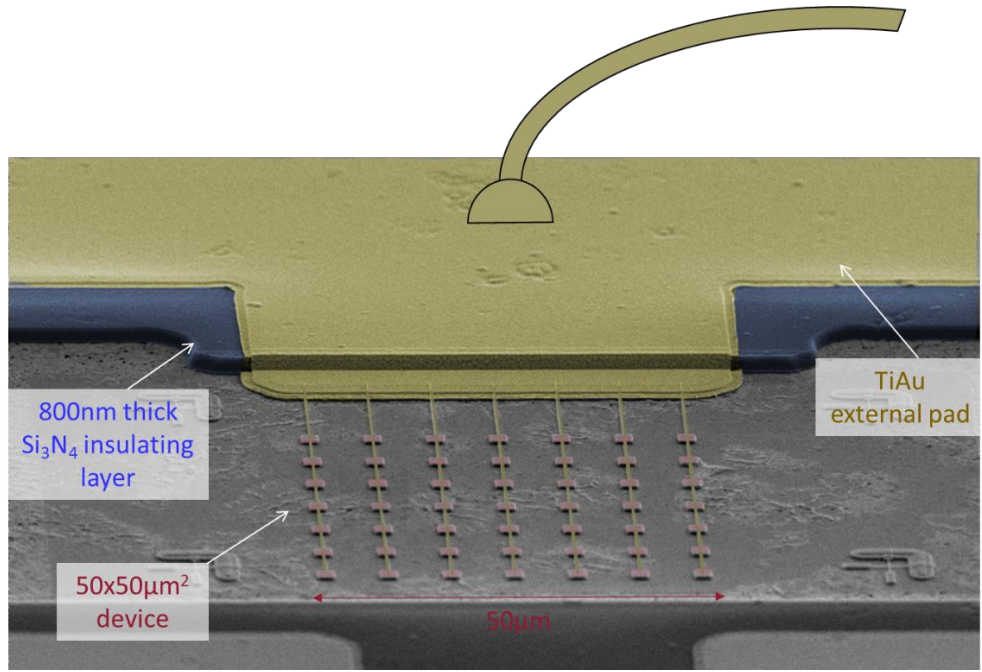

**Fig. S1.** Mid-infrared QWIP structure embedded into an array of patch resonators, pixel size 50μm.

**Light polarization dependence**

Our structures support two fundamental modes, $TM_{100}$ and $TM_{010}$, which are represented in Fig. S2a. This figure shows the vertical electric field $E_z$ in the plane of the resonator, obtained through finite elements simulations. The electric field distribution follows a standing wave pattern, with a node in the center of the square and maxima at the edges. The connecting wires perturb the $TM_{010}$ mode slightly, which results in a lower coupling efficiency for this mode. As a result, the total photoresponse of the antenna-coupled device has a co-sinusoidal dependence with the light polarization of the normally incident wave.

In Fig. S2b, we plot the peak value of the photocurrent for a $s$ = 1.30 μm structure as a function of the polarization of a plane wave incident on the array (open circles), with the 90° direction corresponding to the direction of the connecting wires. The angular integral of Fig. S2b gives a polarization coupling coefficient $\xi_{array} = \int_0^{2\pi} I_{photo}(\theta)d\theta = 71\%$. The product $C\xi_{array}$ quantifies the percentage of incident photons that are gathered by the structure and this allows one to define a photon collection area of each single patch resonator of the array: $A_{coll} = C\xi_{array}p^2$. The contrast value $C$ of the $TM_{100}$ polarized light is obtained from the measurement of Fig. 1b. For comparison, in the same graph we also plot the polarization dependence of the photoresponse measured for the mesa geometry (open squares). Here the 0° direction corresponds to the growth direction of the QWs, and the incident wave propagates normally to the 45° polished facet. This polar plot therefore recovers the inter-subband selection rule, as expected[7,25].

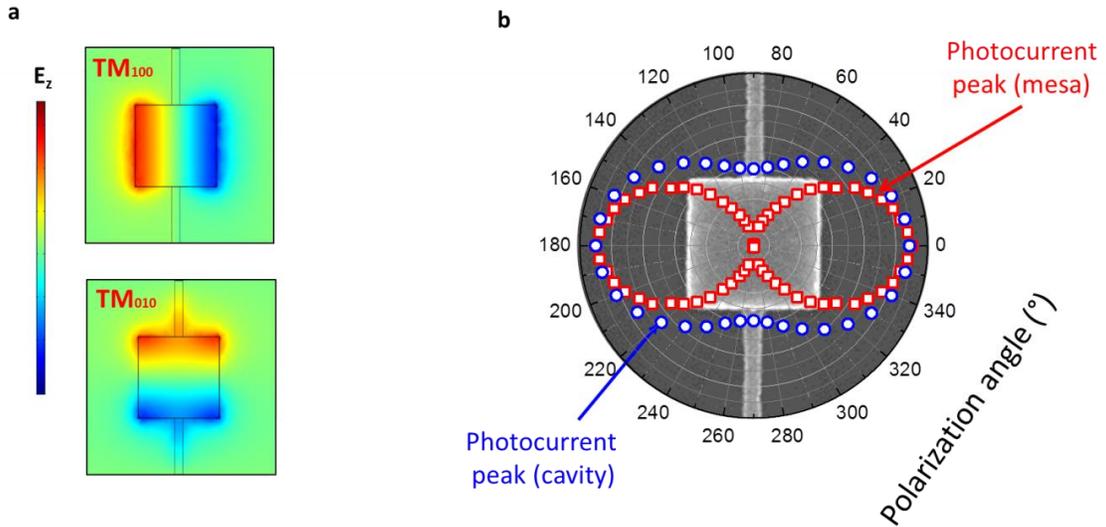

**Fig. S2. a**, Finite element simulation of the $E_z$ field component coupled with the patch cavity QWIP, for the $TM_{100}$ and the $TM_{010}$ modes. **b**, polar graph of the cavity photocurrent peak as function of the wire grid polarization angle. The photocurrent is normalized at its maximum at 0°. The open circles are the results for the cavity array, where the 90° direction corresponds to the connecting wires. The open squares are the results for the mesa geometry, where the 0° direction corresponds to the growth direction of the QWs.

**Responsivity, gain and background-limited detectivity**

In fig. S3a we show the responsivity curves as function of voltage for both the mesa and the patch cavity with s = 1.35 μm. The decrease of the responsivity with temperature is attributed to the thermal dependence of the charge carrier drift velocity and to an increased phonon-electron interaction[14,15] (see Fig. 2c). Note that QWIP devices show the typical negative differential photoconductivity, identified as the Gunn effect, which consists of a photocurrent decrease as function of voltage at specific critical fields, at which inter-valley electron scattering is induced in GaAs[7].

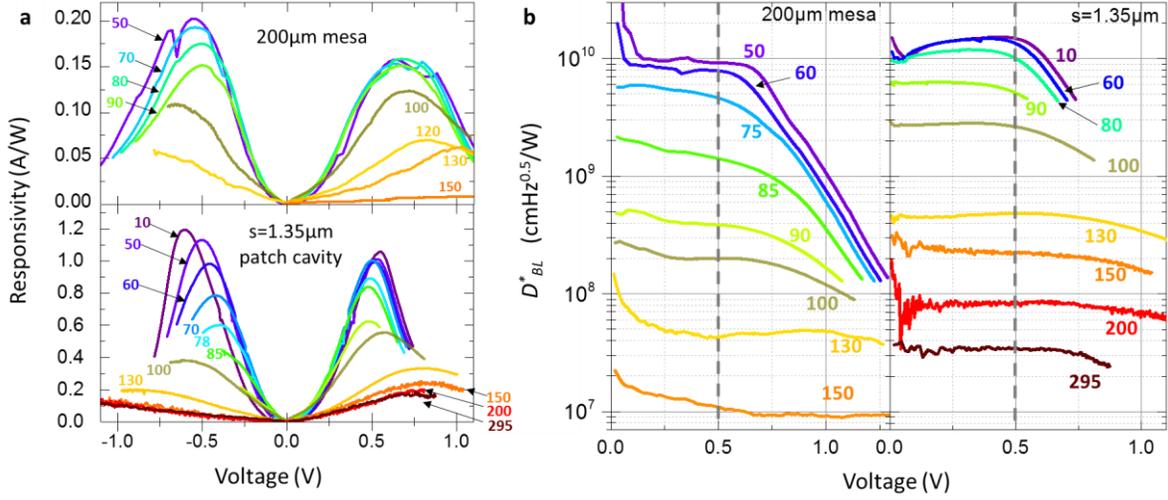

**Fig. S3 a**, responsivity of the mesa and the s=1.35 µm antenna-coupled devices as function of applied voltage. The temperature in K of the QWIP is indicated for each measured curve. **b**, background-limited specific detectivity for the mesa (up to 150 K) and the microcavity (up to room temperature) devices.

The responsivities of the mesa can be expressed by considering the voltage dependent photoconductive gain $g(T,V)$ of the detector active region and the peak inter-subband energy $E_{21}$ = 143 meV (taking into account many-body effects):

$$R_{mesa}(E_{21}, T, V) = \eta_{isb}(E_{21}) \, eg(T,V) \, t_{GaAs} \xi_{mesa}/E_{21} \quad (S1)$$

where $\eta_{isb}$ = 5.0% is the absorption coefficient for the five QW system in the 45° facet geometry, $e$ is the electron charge, $t_{GaAs} = 0.67$ is the substrate transmission coefficient at 8.6 µm and $\xi_{mesa}$ = 0.5 is the polarization factor (only one polarization of the incident light is coupled with the 45° facet). Analogously to Eq. (S1), we can define[21]:

$$R_{array}(E_{21}, T, V) = \frac{B_{isb}(E_{21})}{B_{isb}(E_{21}) + Q_{ohm}^{-1} + Q_{rad}^{-1}} \, eg(T,V) \, C\xi_{array}/E_{21} \quad (S2)$$

where $Q_{ohm}$ = 4 and $Q_{rad}$ = 22 represent the ohmic and radiative dissipation of the double metal cavity, respectively, obtained by reflectivity measurements. The dimensionless parameter $B_{isb}$ quantifies the energy dissipation through inter-subband absorption and is expressed by a lorentzian lineshape:

$$B_{isb}(E) = f_w \frac{E_P^2}{4E_{21}} \frac{\hbar\Gamma}{(E-E_{21})^2 + \frac{(\hbar\Gamma)^2}{4}} \quad (S3)$$

where $f_w = N_{QW}L_{QW}/L=0.067$ is the filling factor of the absorbing QWs on the overall thickness, $E_p$ = 47.2 meV is the inter-subband plasma energy, and $\Gamma$ = 15.0 meV is the full-width-at-half-maximum of the mesa photo-response, obtained by a fit to the experimental data. We obtain a similar value $B_{isb}=0.07$ for the two resonant cavities s = 1.30 µm and s = 1.35 µm. The absorption coefficient in the antenna-coupled QWIPs is described by the branching ratio $\eta_{array} = \frac{B_{isb}}{B_{isb}+Q_{ohm}^{-1}+Q_{rad}^{-1}}$ = 18.9%. Using Eq. (S1) and Eq. (S2) with the measurement data in Fig. **2a**, we obtain very similar values for the photoconductive gain for the mesa and the array, as shown for the data at 0.5 V (21 kV/cm) in Fig. **2a**. This confirms that the absorbing regions for the two geometries are identical. Furthermore, the data shows an exponential decrease of the gain as a function of temperature. Following Ref. **7** the photoconductive gain can be defined as:

$$g = \frac{\tau_{capt} v_d}{N_{QW} L_p} \tag{S4}$$

where $\tau_{capt}$ = 5 ps is the capture time, $v_d$ is the drift velocity, $N_{QW}$ = 5 is the number of quantum wells and $L_p$ = 40.2 nm is the length of a period in the structure. The thermal dependence of the gain is related directly to the drift velocity and therefore to the electron mobility. Following Ref. **14** we can express the temperature dependence as:

$$g(T) = \frac{1}{\frac{1}{g_0} + \frac{B}{\exp\left(\frac{E_{LO}}{k_B T}\right)} + \left(\frac{E_{AC}}{k_B T}\right)^{3/2}} \tag{S5}$$

Here $E_{LO}$=36 meV is the longitudinal optical phonon energy in GaAs, and the fit parameter $g_0$=1.25±0.03 expresses the value of the gain at equilibrium (without thermal scattering dependence). The second term in the denominator represents the polar optical scattering (see Ref. **15**) where the parameter $B$=24.4±1.6 is a dimensionless polar constant and the third term represents the deformation potential scattering caused by interaction of carriers with acoustic phonons, with a corresponding parameter $E_{AC}$=0.07±0.01 meV which characterizes the acoustic deformation potential. Eq. (S5) provides very good fits of the experimental data, confirming the model.

The values of photoconductive gain obtained in this way are used to calculate the background limited detectivity as function of applied voltage, at different temperatures, as illustrated in Fig. S3b.

**Linearity and Heterodyne Measurement**

In Fig. S4 we show the spectra of the two QCLs compared to the room temperature response of the QWIP in the microcavity array geometry. We notice that the lasers are detuned from the maximum inter-subband absorption, resulting in a detector photoresponse that is half of the maximum achievable. This is an important remark because the responsivity and detectivity values we report in Figs. 2 and 3 correspond to the peak values of detector photoresponse. The background-limited NEP (noise equivalent power) is defined as NEP=$\sqrt{A_{det}}$/D*. Using our measured value of detectivity at 295 K for the cavity with s = 1.30 µm at 0.5 V (Fig. 3) we have $D^*$=2.8×10$^7$ cmHz$^{0.5}$/W and NEP = 0.2 nW/Hz$^{0.5}$. Taking into account the 50% spectral overlap, we obtain NEP = 0.4 nW/Hz$^{0.5}$, which agrees with that observed from the linearity measurement in Fig. **4c**. Therefore the data presented in the main text are perfectly consistent.

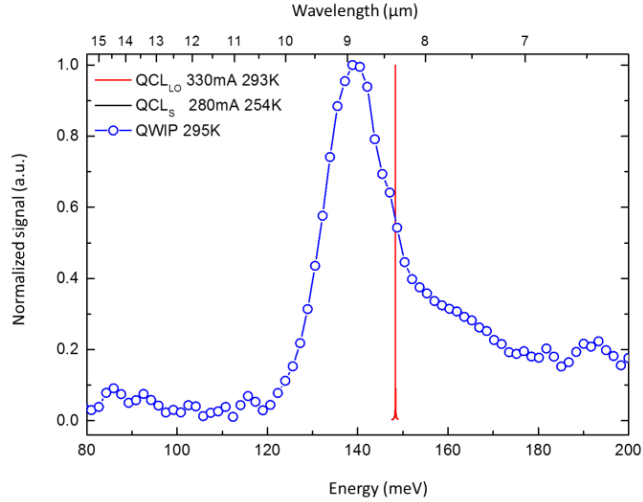

**Fig. S4** Emission spectra of the QC lasers compared to the room temperature response of the microcavity QWIP

For the heterodyne measurement the QC laser used as the LO is kept at a temperature 254 K while the QC laser used for the signal is kept at 293 K. With the temperature stabilized, it is possible to tune the spectral position of the two DFBs by slightly changing the applied DC current, according to the tuning coefficients β$_{LO}$=378 MHz/mA and β$_S$=413 MHz/mA (extracted from a linear fit to the emission frequency of the lasers as a function of temperature and bias).

In the case of a high power LO, the NEP of the heterodyne can be written[7] NEP$_{het}$=E$_{21}$Δf/η where $\eta$ is the absorption coefficient of the QWIP. For our device in the microcavity array we have a theoretical limit of NEP$_{het}$ of less than 1 aW for an integration time of 1 s at 300 K. In the experiment shown in Fig. **4**, the signal-to-noise ratio is still mainly limited by the dark current. The square root fit of the signal-to-noise ratio can be extrapolated to a S/N ratio equal to 1, which provides NEP$_{het}$ ~ 10 fW for an integration time of 1 s (NEP$_{het}$ ~ 1 pW for an integration time of 10 ms), that is still four orders of magnitude higher than the theoretical limit. These estimations indicate that a high power LO could achieve sensitivities at the single photon level at room temperature.